# Performance Characteristics of Mixed Irradiated n-MCz Thin Silicon Microstrip Detector for the HL-LHC Experiments

Anchal Sharma[a], Nitu Saini[a], Shilpa Patyal[a], Balwinder Kaur[a], Ajay K. Srivastava[a,*]

[a]Department of Physics, University Institute of Sciences, Chandigarh University,
Gharuan-Mohali, Punjab,140413, India.

**Abstract**

Within CERN RD50 Collaboration, MCz Si was identified as a best material for the detector used in the high luminosity collider experiments. The n in p Si detectors were utmost radiation hard detectors, which can be used for the phase 2 upgrade plan of the new Compact Muon Solenoid tracker detector in 2026. The choice of n or p-MCz Si material depends upon the high electrical charge collection efficiency performance of these detectors over n or p-FZSi detectors. The bulk radiation damage model for n, p-MCz Si need to develop for the designing, development and optimization of the advanced detectors for the next generation High Energy Physics experiments.

In this work, an advanced four level deep-trap mixed irradiation model (E5, $C_iO_i$, E(30K), H(152K)) for n-MCz Si is proposed by the comparison of experimental data on the full depletion voltage and leakage current to the Shockley Read Hall recombination statistics results on the mixed irradiated n-MCz Si PAD detector. Prediction uncertainty in the radiation damage mixed irradiation model considered in the full depletion voltage and leakage current, which was due to the uncertainty in macroscopic results from an experimental measurement on the mixed irradiated n-MCz Si pad detector. A very good agreement is observed in the experimental and SRH results on the full depletion voltage and leakage current. This model is also used to extrapolate the value of the full depletion voltage at different mixed (proton + neutron) higher irradiation fluences for the thin n-MCz Si microstrip detector.

*Keywords*:Si microstrip detector; TCAD simulation; Bulk damage; Full depletion voltage;Leakage current;CCE; Mixed irradiation.

## 1. Introduction

The Large Hadron Collider(LHC) will be upgraded to theHigh-Luminosity Large Hadron Collider(HL-LHC) to increase the potential of the experiment for the new discoveries in physics in 2026 [1]. The extremely high-integrated luminosity 3000 $fb^{-1}$ demands the high-performance radiation hard advanced Si detectors to survive in the hostile mixed irradiation environment at the HL-LHC experiment. The current Compact Muon Solenoid (CMS) tracker system of the CMS detector is not capable of particle tracking, which was due to occurrence of high Pile-up events. Therefore, the present CMS tracker system must be replaced with a new high radiation tolerance detector in order to provide the proper particle tracking for each bunch crossing in the proton-proton collisions. Therefore,thin advanced n in p radiation hard Si strip detector is required for the outer tracker of the new CMS detector at HL-LHC [2-9].

[*] Corresponding author. Tel.: +91-8400622542; e-mail: kumar.uis@cumail.in

The choice of the material of this advanced n in p Si detectors was n or p-MCz (Magnetic Czochralski) Si. The electrical performances of the detectors were needed to test for the designing, development and optimization of the detectors for the experimental HEP experiments. Within CERN RD50 Collaboration for the HL-LHC upgrade plan, n/p FZ (Float zone) Si detectors and n/p MCz Si pad detectors were exposed to protons, neutrons and mixed (proton +neutron) irradiation. The macroscopic and microscopic performance of these detectors measured using current-voltage (I/V), capacitance-voltage (C/V), Thermally Stimulated Current (TSC), Deep Level Transient spectroscopy (DLTS), Transient Current Technique (TCT) and Alibaba system SL., Barcelona, Spain set up. Several Italian, German and Indian groups have proposed the p-FZ Si radiation damage models for the radiation damage analysis of the irradiated Si detectors.

In this paper, we have compared the macroscopic results on the full depletion voltage and leakage current in n-MCz Si pad detector at different fluences [5] to the SRH (Shockley Read-Hall) calculations. Finally, we have to tune the capture cross-sections values of electrons and holes within 5% of its value and introduction rates of E30K in order to obtain the good agreement in the SRH and experimental data. In this paper the advanced four level deep-trap radiation damage model (E5, $C_iO_i$, E(30K), H(152K)) is proposed for the n-MCz mixed irradiated detector.

The model is useful to explain the macroscopic performance of the n-MCz Si PAD detectors in the mixed irradiation environment. This model can be fed into the Technology Computer-Aided Design(TCAD) device simulation programs for the radiation damage analysis of the mixed irradiated Si strip detector and the performance characteristics can be beautifully explained using the electric-field distributions and space charge distributions.

## 2. Performance characteristics of the Si detector

The electrical performance of the n or p-MCz Si detectors than FZ Si detectors shows remarkable results in Full Depletion Voltage ($V_{fd}$) and Charge Collection Efficiency (CCE) [5] that makes MCz an intrinsic choice as a detector material for the development of the advanced radiation strip detectors for the next generation High Energy

Physics (HEP) experiments. Fig. 1 shows the FLUKA simulation of the fluence levels in the CMS Tracker after 3000 fb$^{-1}$.

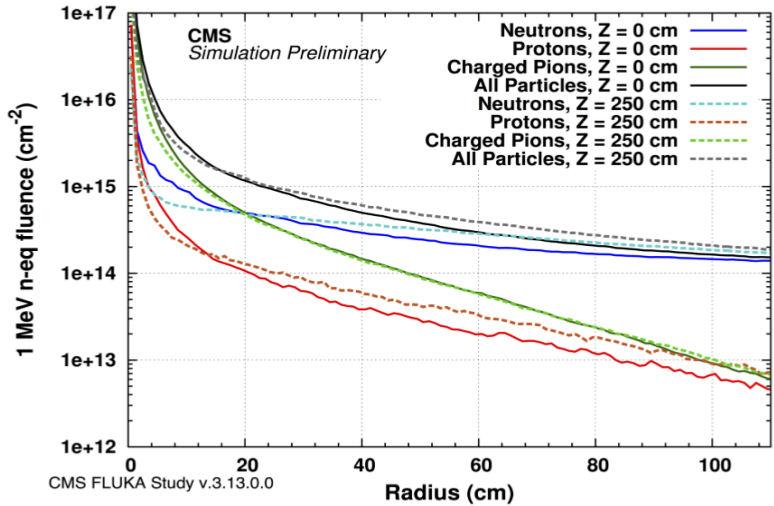

Fig. 1. The FLUKA simulation of the fluence levels in the CMS Tracker after 3000 fb$^{-1}$. The Z-coordinate in the legend refers to the distance from the interaction point along the beam line [4].

*2.1. Effect of mixed irradiation on n/p FZ and n/p MCz Si pad detector*

For both n/p FZ Si pad detectors, damage is accumulated in n or p-FZ Si, weather detector is exposed to proton or mixed irradiation. From [5] it is clear that $V_{fd}$ increases with the fluences up to $5 \times 10^{14} n_{eq.}/cm^2$ for both n or p-FZ Si. As expected, $V_{fd}$ will increases and it can be more than 1000 V for both n/p FZ Si detectors irradiated with proton or mixed irradiation for fluence $9 \times 10^{14} n_{eq.}/cm^2$. Therefore, FZ Si material cannot be used in the phase 2 upgrade of the new CMS tracker for the HL-LHC experiments.

For n-MCz detector, damage is compensated in case of mixed irradiation compared to proton irradiation because donor deep trap defect E(30K) introduced in proton irradiation is balanced by the acceptor H(152K) deep trap defect introduced in the neutron irradiation (see table 2). Therefore, both of these trap defects balance each other and thus $V_{fd}$ decreases in mixed irradiated n-MCz Si pad detector. Hence, n-MCz can be considered as a prime candidate for the mixed irradiation fluence of the new CMS tracker upgrade plan.

In case of mixed irradiated p-MCz Si pad detector the damage is not compensated. $V_{fd}$ is higher in mixed irradiated p-MCz Si detector than proton irradiated detector [5]. But $V_{fd}$ is less in mixed irradiated p-MCz at higher fluence than n-MCz Si pad detector therefore, p-MCz can also be an option for the new CMS detector. Ajay K. Srivastava detector group in India is already working on mixed irradiated p-MCz detectors for the next generation HEP experiments.

**3. Advanced four level deep trap mixed irradiation model for n-MCz Si pad detector**

In this section, we have discussed the device model and SRH methods that can be used to develop the advanced four level deep-trap mixed irradiation model for n-MCz Si pad detector.

*3.1. Device model and SRH calculations*

A rectangular cell of 0.0625 cm$^2$ x 300 μm n-MCz Si pad detector model is used for the SRH calculations (see Fig.2) and the device and process parameters of the detector are shown in Table 1. Fig. 2 shows the cross-section of the n-MCz Si pad detector irradiated with mixed irradiations used in the experimental work [5]. The detector is having Dirichlet and Neumann boundary conditions on the respective electrodes and free open boundary surface.

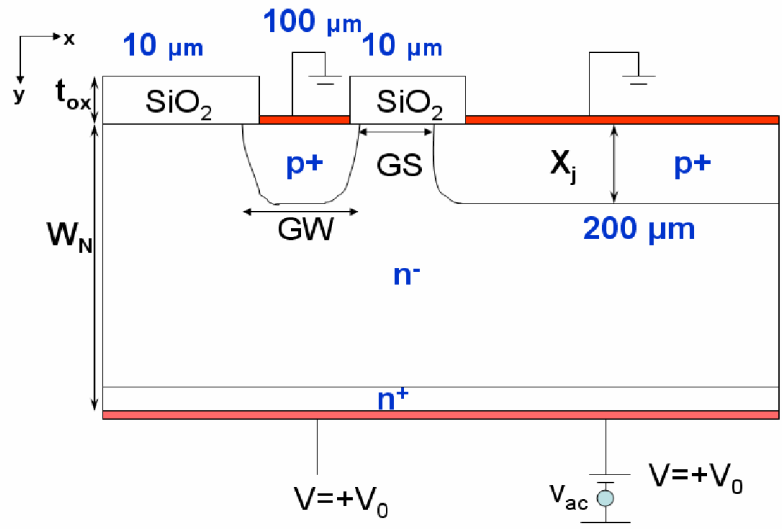

Fig. 2. Cross-section of the 0.0625 cm$^2$ x 300μm n-MCz Si pad detector model used in the present study for SRH calculations.

Table 1. Device and process parameters of n-MCz Si pad detector for mixed irradiation.

| S.No. | Physical parameters | Values |
|---|---|---|
| 1. | Doping concentration ($N_D$) | $2.87 \times 10^{12}$ cm$^{-3}$ |
| 2. | Oxide thickness ($t_{ox}$) | 0.5 μm |
| 3. | Junction Depth ($X_j$) | 1 μm |
| 4. | Guard ring spacing | 10 μm |
| 5. | Guard ring width | 100 μm |
| 6. | Device depth | 300 μm |
| 7. | Fixed oxide charge ($Q_f$) | $1 \times 10^{12}$ cm$^{-2}$ |
| 8. | Resistivity | 1.5 KΩ-cm |

*3.1.1. Comparison of $V_{fd}$ (experimental) with theoretical $V_{fd}$ (SRH)*

The detector used as a model for the SRH calculations (see Fig.2 and Table.1). By using theoretical SRH equations [6], we have estimated SRH $V_{fd}$ and compared this with the experimental value of $V_{fd}$. A very good agreement has been observed in both the $V_{fd}$ values (see Fig.3). The uncertainties in the value of $V_{fd}$ for the theoretical SRH calculations can be up to 10%, for the different mixed irradiation fluences. This uncertainty in extracted $V_{fd}$ from the ln C-ln V plot can be also observed in an irradiated MCz Si PAD detector. For the accurate estimation of the $V_{fd}$(experimental) in the experimental situations, an uncertainty of 10% is taken into account in all calculations. Table.2 shows the advanced mixed irradiation "four level deep-trap model" for n-MCz Si pad detector.

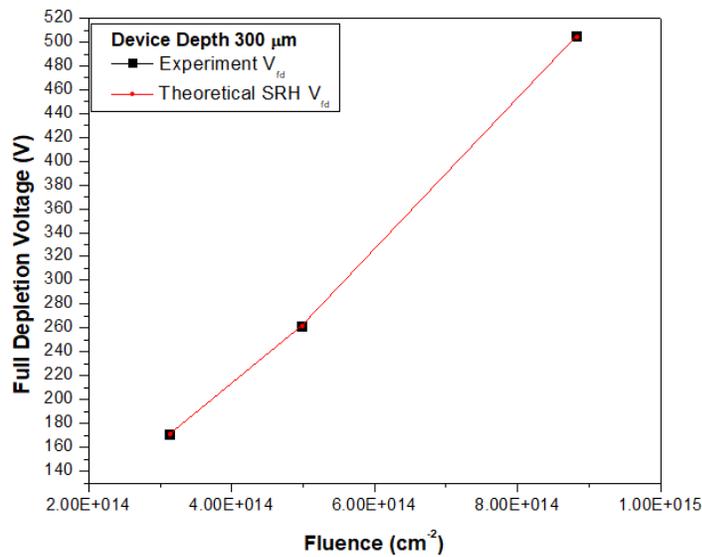

Fig. 3. Comparison of experimental [5] and theoretical SRH value of $V_{fd}$.

Table 2. Advance Mixed irradiation "four level deep traps model" for n-MCz Si pad detector.

| Defect/type | Effects on the macroscopic parameters | Energy level (eV) | $\sigma_n$ [cm$^{-2}$] | $\sigma_p$ [cm$^{-2}$] | $\eta$ [cm$^{-1}$] |
|---|---|---|---|---|---|
| E5 / Acceptor | Increase of leakage current | $E_C$-0.46 | 1.41x10$^{-15}$ | 2.79 x10$^{-15}$ | 12.4 |
| H (152K) / Acceptor | -ve space charge | $E_V$+ 0.42 | 4.58 x10$^{-13}$ | 6.15 x10$^{-13}$ | 0.04 |
| $C_iO_i$ / Donor | +ve space charge | $E_V$+ 0.36 | 2.08 x10$^{-18}$ | 2.45 x10$^{-15}$ | 1.1 |
| E (30K)/ Donor | +ve space charge | $E_C$-0.10 | 2.30 x10$^{-14}$ | 2.00 x10$^{-15}$ | See Fig. 5 |

As a result, a new and advanced four-level deep trap model for mixed irradiation is proposed, which is shown in Table 2, where E5, $C_iO_i$, E(30K), H(152K) are the four deep trap defects. E5 and H(152K) are acceptor defect and $C_iO_i$ and E(30K) are donor trap defects. The effective capture cross-section of the n-MCz deep traps and introduction rates are tune in order to get the good agreement with the experimental data [7]. Here, in the case of n-MCz mixed irradiated Si pad detector, the effective introduction rate of E(30K) in n-MCz or in p-MCz Si plays an important role and that can be a key trap to explain the macroscopic performance of the n/p-MCz pad detector.

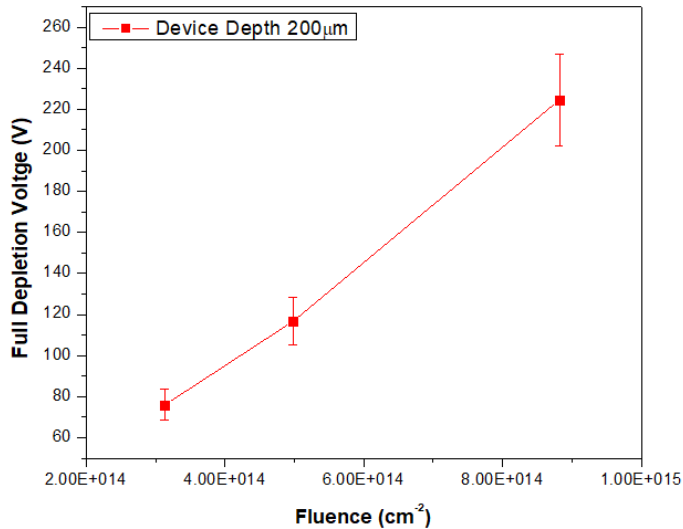

Fig. 4. Full depletion voltage for 200μm thin detector.

Now, it was interesting to see the $V_{fd}$ behaviour for the thin MCz-Si pad detector, and compared it with the experimental $V_{fd}$ value of 300μm thick n-MCz Si pad detector (see Fig. 3). From Fig. 3, 4, we concluded that the $V_{fd}$ increases with the mixed irradiation fluences for both thick and thin detector. The $V_{fd}$ value for thin mixed irradiated detector is very less (< 200 Volt) as compared to $V_{fd}$ value for thick mixed irradiated detector for the same fluences, which is acceptable $V_{fd}$ range as per the specifications of the irradiated detectors at HL-LHC.

*3.1.2. Introduction rate of E(30K) in mixed irradiated environment*

The introduction rate of E(30K) is not constant as per other deep traps and increases slightly with an increase in the mixed irradiation fluences, which is shown in Fig. 5.

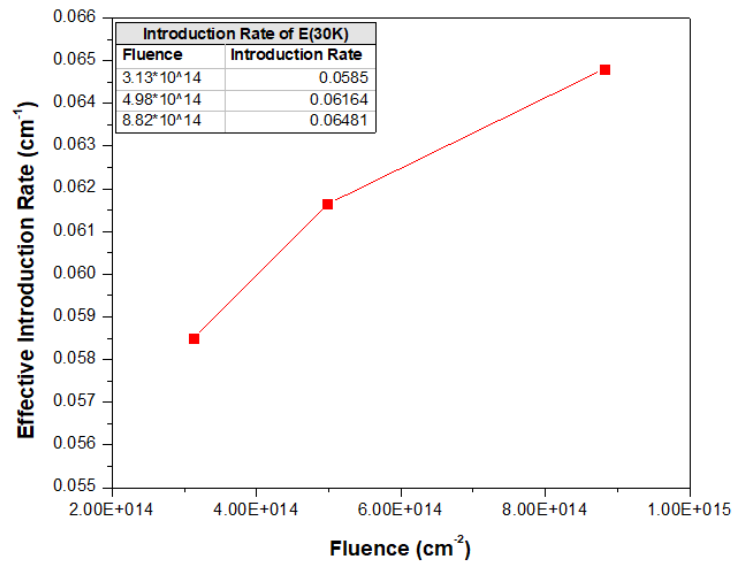

Fig.5. Effective Introduction rate of E (30K) as a function of the mixed irradiation fluences.

*3.1.3. Leakage current in mixed irradiated n-MCz Si pad detector*

The generation leakage current is not a major issue in many radiation damage detection systems even it increases the overall noise of the detector system. Using multiple field guard rings structure on outer surface of the detector and by cooling the detector system in the experiment up to $-20^0C$ to $-30^0C$ we can control the increased leakage current. In the experiment [5], temperature was measured few cm away from the sample which was stable within ±1°C that result in 15-20% error in leakage current measurement. Here, we have estimated the leakage current by SRH calculations and then compared it with the experimental value at 293K. It is noted that a very good agreement has been observed in the experimental and theoretical leakage current at 297K (Exp. Temperature+4K), which is shown in Fig.6. Experimental (293K) and SRH (297K) leakage current calculations differs by a factor of 1.02. A less estimation of the SRH leakage current is observed at 293K than experimental leakage current, which may be also due to the absence of the physical models; Impact ionization, Hurkx Trap Assisted Tunnelling, interface oxide and many more in the SRH calculation, and a high-field as expected in the mixed irradiated real n-MCz Si pad detector. Therefore, we can say that this advanced n-MCz four level deep traps model is in good agreement with the experimental data on the mixed irradiated Si detectors. The comparison of experimental and theoretical SRH data on the leakage current with 10% uncertainty is shown below in Fig. 6.

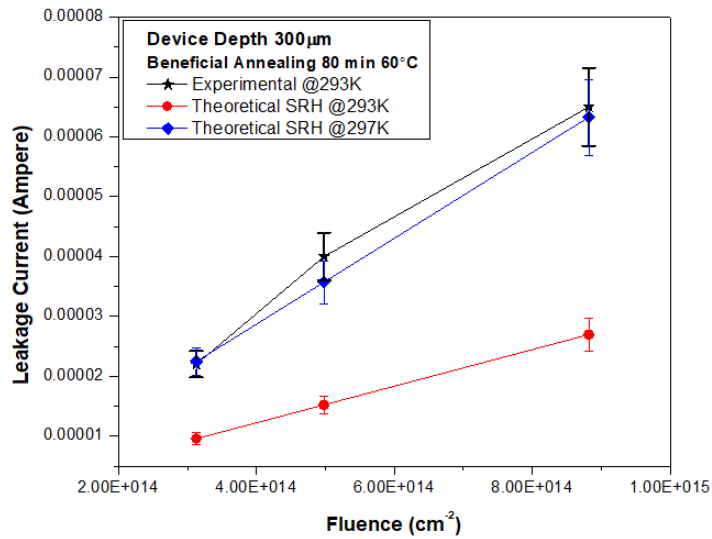

Fig. 6. Leakage current as function of the mixed irradiated fluence at 293 K (beneficial annealing of 80 min $60^0$C).

### 3.1.4. Extrapolated value of $V_{fd}$ in higher mixed irradiation environment

We have extrapolated the $V_{fd}$ value at higher mixed irradiation fluences for the thick and thin n-MCz Si pad here, consider strip detector (See Table.1 for doping concentration). Due to the safety margin of the detector operation in the hostile radiation environment of the HL-LHC during the irradiation of detector, we have taken the higher mixed irradiation fluences of the order 4.50 x $10^{15}$ $cm^{-2}$. We have plotted the extrapolated $V_{fd}$ value as a function of the mixed irradiation fluence for the n-MCz Si pad detector (see table 1,2 and Fig. 2) and the $V_{fd}$ results are shown in Fig. 7.

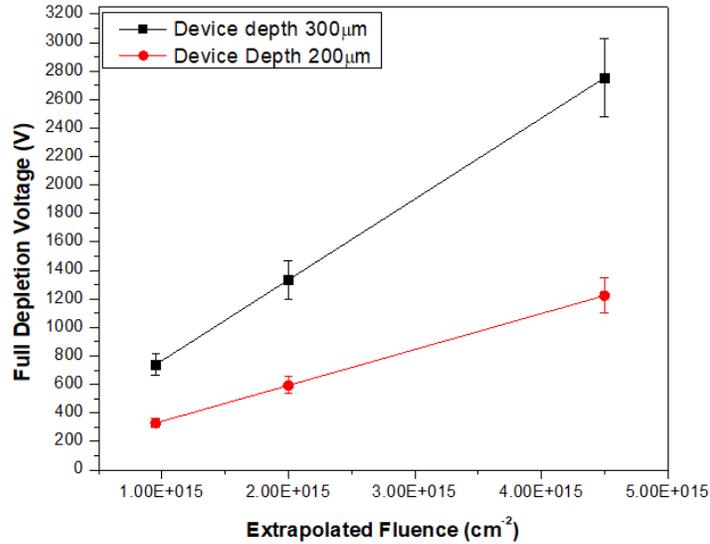

Fig. 7. Extrapolated value of $V_{fd}$ for thick (300μm) and thin (200μm) mixed irradiated n-MCz Si detector.

The extrapolated value of $V_{fd}$ is nearly 1200V at fluence $4.5 \times 10^{15}$ for the 200μm thin mixed irradiated detector whereas, for thick detector 300μm, $V_{fd}$ was too high as expected at higher fluences.

Finally, we proposed an advanced four level deep trap mixed irradiation model for the n-MCz Si and that can be used to design and optimized the radiation hard n-MCz thin Si strip detector for HL-LHC experiments.

**Conclusion**

In this present work, we have proposed an advanced four level deep-trap mixed irradiation model for the n-MCz by comparing the results of SRH theoretical calculations and experimental results on the mixed irradiated n-MCz Si pad detector.

A very good agreement has been observed in the SRH theoretical and experimental data on the full depletion voltage and leakage current at (experimental temperature +4K) 297 K for the mixed irradiated detector. The uncertainty in $V_{fd}$ and leakage current is taken account for the better understanding of the experimental results. Using our radiation damage model, the extracted $V_{fd}$ of thin n-MCz Si strip detector irradiated by mixed irradiation fluence of $4.5 \times 10^{15}$ cm$^{-2}$ is around 1200 V. Therefore, radiation hard thin n-MCz Si microstrip detector for the new CMS tracker detector system can be designed and optimized, which can work up to 1500V at HL-LHC experiments. The CCE behaviour is extremely important measure for the performance of the n/p-MCz detector system at HL-LHC.